\documentclass[10pt,twoside,onecolumn]{article}
\usepackage[]{latexsym,amsmath,amssymb}
\usepackage[]{epsfig}
\newcommand{\be}{\begin{eqnarray}}
\newcommand{\ee}{\end{eqnarray}}
\newcommand{\ud}{\mathrm{d}}

\newcommand{\lp}{\ell_{\rm p}}
\newcommand{\mpl}{M_{\rm p}}

\title{\bf The decay-time of non-commutative micro-black holes}
\author{Roberto Casadio$^a$\thanks{casadio@bo.infn.it}$\ $
and
Piero Nicolini$^b$\thanks{piero@cmfd.units.it}
\\
\\
$^a$Department of Physics, University of Bologna,
and I.N.F.N.,
\\
via~Irnerio~46, 40126~Bologna, Italy
\\
$^b$Physics Department,
California State University Fresno,
\\
Fresno, California 93740-8031 and
\\
Department of Mathematics and Computer Science,
\\
Consortium of Magneto-Fluid-Dynamics,
University of Trieste and I.N.F.N.,
\\
via~Valerio~12, 34127~Trieste, Italy}
\begin{document}
\maketitle
\begin{abstract}
Non-commutative black holes are characterised by a minimum mass which would
result in a remnant after the Hawking evaporation ends.
We numerically study the decay of neutral non-commutative black holes
for up to ten spatial dimensions and typical parameters that would make their
production possible at the LHC.
Neglecting possible accretion mechanism, we find that decay-times are extremely short.
\end{abstract}
\section{Introduction}
\setcounter{equation}{0}
One of the most striking predictions of General Relativity is the existence of black holes,
that is event horizons which surround and hide the point-like or otherwise singular
classical sources.
On general quantum-mechanical arguments, one however expects that singular
sources are smeared out and there are indeed many theoretical reasons to believe the
fabric of space-time can be effectively described by non-commutative geometry at short
scales (see, for example Section~2 of Ref.~\cite{nicolini} and Ref.~\cite{szabo}).
Such modifications to General Relativity must necessarily have an impact on the structure
of black holes.
\par
The non-commutative analogue of the Schwarzschild black hole metric was found
in Ref.~\cite{nicolini06} (see also Ref.~\cite{nicolini05}) and subsequently
generalised to include the electric charge~\cite{nicolini07} and extra-spatial
dimensions~\cite{rizzo06,nicolini08}.
A common feature for all such metrics is that there exists a minimum mass for
the black hole which solely depends on the non-commutative length
$\ell$~\footnote{This length is related to the non-commuative parameter $\theta$
of Ref.~\cite{nicolini} by $\ell=2\,\sqrt{\theta}$.}
and the fundamental scale of gravity $M_g=\ell_g^{-1}$ (for a comprehensive and
updated review, see Ref.~\cite{nicolini}).
The former is usually identified with the Planck length $\lp$ and the latter with the Planck
mass $\mpl$ in four space-time dimensions, but $\ell$ could actually be much larger and
$M_g$ much lower, possibly around $1\,$TeV, if extra-dimensions exists~\cite{add,rs}.
This has opened up the possibility of having micro-black holes (mBH) with a mass of a few
TeV's~\cite{banks} that might be produced and detected
at the Large Hadron Collider (LHC)~\cite{dimopoulos,ch02,mbh}.
\par
The standard picture of mBH at the LHC is that they should be produced with a mass
of around $10\,$TeV and very rapidly decay completely~\cite{dimopoulos} via the Hawking
effect~\cite{hawking}.
Generally small (but possibly significant) corrections to this scenario arise from the
use of the microcanonical picture~\cite{mfd}, which essentially enforces energy
conservation throughout the evaporation process and suppresses the emission
for small mBH mass~\cite{ch02,c07}.
If the non-commutative description of mBH given in Ref.~\cite{nicolini} is correct,
one however expects that a (stable) remnant will be left at the end of the Hawking
evaporation~\cite{hosse}, which might exit the accelerator and detectors.
In this context, and for phenomenological purposes, it is therefore important
to estimate the typical decay-times for such objects.
For example, the issue of mBH life-times has recently been debated
in Refs.~\cite{mangano1}.
\par
Since it is widely accepted that mBH discharge very quickly, via several processes,
including the Schwinger mechanism at least for $3\le d\le 5$~\cite{nicolini08},
we shall here consider only the neutral metrics of Refs.~\cite{nicolini,rizzo06}
for $d=5,\ldots,10$ spatial dimensions (noting that higher values of $d$ seem to be
favoured~\cite{d>4}).
In order to have mBH produced at the LHC, we need a minimum mBH mass of
order $1\,$TeV, which in turn implies that $\ell_g\sim \ell$.
Further, we shall see that the typical temperature of such mBH's always remains
smaller than their mass, and microcanonical corrections can thus be neglected.
The evolution of the mBH mass is then determined by solving the relevant equation
numerically for initial mBH mass of order $10\,$TeV and an upper bound for
the decay-time obtained.
\par
We shall use units with $c=\hbar=1$.
\section{Non-commutative neutral mBH}
\setcounter{equation}{0}
\label{mBH}
We recall here that the non-commutative metric of a neutral black hole of
(asymptotic) proper mass $m$ in $d+1$ space-time dimensions is given
by~\cite{nicolini08}
\be
\ud s^2=-A_d\,\ud t^2+A_d^{-1}\,\ud r^2+r^{d-1}\,\ud\Omega^2_{d-1}
\ ,
\label{metric}
\ee
where
\be
A_d=1 -\frac{2\,G_d\,m }{r^{d-2}\,\Gamma(d/2)} \,
\gamma\left(\frac{d}{2};\frac{r^2}{\ell^2}\right)
\ ,
\ee
and
\be
\gamma(a;x)=\int_0^x u^{a-1}\,e^{-u}\,\ud u
\ .
\ee
With $d=3$ spatial dimensions, the Newton constant $G_3=G_{\rm N}=\lp/\mpl=\lp^2$
and one usually assumes that $\ell\simeq \lp$.
In the presence of extra spatial dimensions, $d>3$, we can likewise write
$G_d=\ell_g^{d-2}/M_g=\ell_g^{d-1}$, but keeping $\ell$ and $\ell_g$ distinct.
Of course, the non-commutative length must be short enough to agree with present
experimental bounds on the validity of the Newton law at short distance, that is
$\ell\lesssim 1\,\mu$m~\cite{hoyle} and for the gravitational mass we
shall assume $M_g\simeq 1\,$TeV.
\par
The radial mass-energy function for the above solution is given by
\be
m(r)=\frac{m}{\Gamma(d/2)}\,\gamma\left(\frac{d}{2};\frac{r^2}{\ell^2}\right)
\ ,
\ee
and
\be \lim_{r\to\infty} m(r)= m
\ ,
\ee
as anticipated.
Further, for sufficiently large proper mass, there exists two horizons, the outer
one being located at $r=r_+$, which is related to $m$ by
\be
M_d=\ell_g^{d-2}\,\frac{m}{M_g}
=r_+^{d-2} \frac{\Gamma\left({d/2}\right)}{2\,\gamma\left(\frac{d}{2};\frac{r^2_+}{\ell^2}\right)}
\ ,
\label{Madm}
\ee
where $\Gamma(a)$ denotes the Gamma function and $M_d=G_d\,m$ the
Arnowitt-Deser-Misner (ADM) mass of the black hole~\footnote{Note that $M_d$ has
dimensions of (length)$^{d-2}$.}.
The inverse Hawking temperature (in geometrical units) is finally given by
\be
\beta_d= 4\,\pi\, r_+
\left[(d-2)-2\,\left(\frac{r_+}{\ell}\right)^{d}
\frac{e^{- r_+^2/\ell^2}}{\gamma(\frac{d}{2};\frac{r_+^2}{\ell^2})}\right]^{-1}
\ .
\ee
\section{Hawking evaporation}
\setcounter{equation}{0}
\label{dM}
In order to study the evaporation of a mBH, we need to estimate the corresponding
Hawking flux $\Phi$, that is the amount of energy emitted through a unit area per
unit time in the form of thermal radiation.
The latter is described by quantum fields propagating on the $(d+1)$-dimensional
background~(\ref{metric}) and we recall that the space-time non-commutativity
makes the field theory UV~finite, thanks to the presence of a damping term in the
momentum-space propagator,
namely $G(p)\sim\exp\left(-\ell^2\,p^2/8\right)$ (for the details, see Ref.~\cite{Smailagic04}).
As a result, considering for simplicity the case of a massless scalar field  and neglecting the
grey-body factor, one finds
\be
\Phi=2\int \frac{\ud^{d}p}{(2\,\pi)^d}\,
\frac{e^{-\frac{1}{8}\,\ell^2\, p^2}\,p}{e^{\beta_d\,p}-1}
\ .
\ee
From this expression, one can then obtain the total luminosity $\mathcal{L}_d$
which governs the time dependence of the mBH proper mass in the canonical
picture~\cite{nicolini08},
\be
-\frac{\ud m}{\ud t} \propto F_d(r_+)\,
\sum_{n=0}^\infty \frac{(-1)^n}{n!}\left(\frac{\ell}{2\,\beta_d}\right)^{2n}
\Gamma(2n+4)\, \zeta(2n+4)
\equiv
\mathcal{L}_d(r_+)
\ ,
\label{dMdt}
\ee
where $\zeta(a)$ denotes the Riemann zeta function.
\par
The exact form of the function $F_d$ in the above expression is actually difficult to
determine and for a number of reasons.
To begin with, one should consider the contribution of all the different kinds of particles
that can be emitted, like Standard Model fermions and bosons as well as (bulk and brane)
gravitons.
Further, particles with different spins propagate differently and their emission is thus
suppressed by spin-dependent grey-body factors,
which, in turn, strongly depend on the precise near-horizon geometry.
In this respect, there are arguments which suggest that Standard Model particles propagating
only in our $(3+1)$-dimensional space-time makes for most of the Hawking
radiation~\cite{emparan}, so that $F_d\propto r_+^2$ (the area of the brane-section
of the horizon), as well as arguments against this scenario~\cite{cardoso},
according to which one should instead have $F_d\propto r_+^{d-1}$
(the area of the bulk horizon, neglecting a possible squeezing~\cite{mazza}).
Finally, one realistically expects that mBH at the LHC would be produced with an
intrinsic spin (see, for example, Ref.~\cite{flachi}), but no rotating analogue of the
metric~\eqref{metric} is known to date.
\par
Since our main goal here is to provide order of magnitude estimates of the
typical decay-times, we shall consider three possibilities:
\begin{description}
\item[B)]
The mBH emits all the particles in the entire bulk:
\be
F_d^{\rm B}=r_+^{d-1}\,\beta_d^{-(d+1)}
\ .
\label{FdB}
\ee
\item[b)]
The mBH emits like a four-dimensional black hole except for the modified
(inverse) temperature:
\be
F_d^{\rm b}=r_+^2\,\beta_d^{-4}
\ .
\label{Fdb}
\ee
\item[bB)]
The mBH evaporates with a dependence on the temperature like that in Eq.~\eqref{FdB}
and on the 2-dimensional horizon area like in Eq.~\eqref{Fdb}:
\be
F_d^{\rm bB}=\ell_g^{d-3}\,r_+^{2}\,\beta_d^{-(d+1)}
\ .
\label{FdbB}
\ee
\end{description}
All the above forms will be used for the numerical simulations reported on later.
\par
In order to solve~\eqref{dMdt} numerically, we introduce dimensionless quantities
(denoted by a bar) by expressing everything in units of $\ell$ (to a suitable power).
For example, the dimensionless ADM mass will be given by
\be
\bar M_d=M_d\,\ell^{2-d}
\ ,
\ee
and the proper mass by
\be
\bar m=m\,\ell=\bar M_d \left(\frac{\ell}{\ell_g}\right)^{d-1}
\ .
\ee
We shall sometimes find it more convenient to express the proper mass $m$
in units of $M_g$ (denoted by a tilde) as
\be
\tilde m=\frac{m}{M_g} =\bar M_d \left(\frac{\ell}{\ell_g}\right)^{d-2}
\ .
\label{m}
\ee
Since we assume $M_g\simeq 1\,$TeV, $\tilde m$ will just be the mBH mass in TeV's.
Next, we shall need to truncate the series in Eq.~\eqref{dMdt} to a maximum integer
value $n=n_{\rm max}$.
With the above assumptions and redefinitions, Eq.~\eqref{dMdt} becomes
\be
\frac{\ud \bar m}{\ud \bar t}
=
-C\,F_d(\bar r_+)\,
\sum_{n=0}^{n_{\rm max}} \frac{(-1)^n}{4^n\,n!}\,\bar\beta_d^{-2n}
\,\Gamma(2n+4)\, \zeta(2n+4) \equiv -\bar{\mathcal{L}}_d(\bar r_+)
\ ,
\label{dMdtx}
\ee
where $C$ is a numerical constant which we shall comment upon later.
Finally, by making use of Eq.~\eqref{Madm}, we can obtain an equation which
contains the horizon radius as the only time-dependent variable,
\be
\frac{\ud \bar r_+}{\ud \bar t}
=
-\frac{\ud \bar r_+}{\ud\bar m}\,\bar{\mathcal{L}}_d(\bar r_+)
=
-\frac{\bar \beta_d\,\gamma\left(\frac{d}{2};\bar r_+^2\right)}
{2\,\pi\,\bar r_+^{d-2}\,\Gamma(d/2)} \left(\frac{\ell_g}{\ell}\right)^{d-1}
\bar{\mathcal{L}}_d(\bar r_+)
\ ,
\label{drdtx}
\ee
which we now proceed to study numerically, with the aim of determining the
evolution of a mBH with initial mass $m(t=0)\simeq 10\,$TeV.
\section{Numerical results}
\setcounter{equation}{0}
\label{num}
Upon employing the three forms listed in the previous Section for the function
$F_d$, we always found the same qualitative picture.
In particular, for fixed $C$, the choice of $F_d^{\rm B}$ in Eq.~\eqref{FdB} always
produced the longest decay-times for $5\le d\le 10$.
Moreover, for a given initial mBH mass and choice of $F_d$, the decay-time
increases for increasing $d$ and is (roughly) linearly proportional to $C$.
Since larger values of $d$ are anyway favoured~\cite{d>4}, we shall therefore
show the detailed analysis just for the case $d=10$ and $F_{10}=F_{10}^{\rm B}$
with $C=1$.
This choice yields the longest decay-time overall (for fixed $C=1$),
which can thus serve as a ``worst'' (or ``best'') case scenario, and we shall further
comment about other cases when relevant.
\par
First of all, we need the minimum mBH mass.
This is obtained by first minimising the ADM mass~\eqref{Madm} with respect to
$r_+$, which, for $d=10$, yields the minimum (dimensionless) horizon radius
$\bar r_{\rm min}\simeq 1.11$, corresponding to a minimum ADM mass
$\bar M_{\rm min}\simeq 133$.
In order for the latter to translate into a minimum proper mass
$m_{\rm min}\simeq 1.2\,$TeV, we thus set (see Eq.~\eqref{m})
\be
\ell_g\simeq 1.8\,\ell
\ .
\label{lg}
\ee
For smaller values of $d$, one obtains slightly smaller $\ell_g$'s, but with a very weak
dependence on the space dimension.
In fact, for $d=5$, one has the smallest value $\ell_g\simeq 1.4\,\ell$.
\par
\begin{figure}[t]
\centering
\raisebox{3cm}{$\bar\beta_{10}^{-1}$}
\epsfxsize=6.5cm
\epsfbox{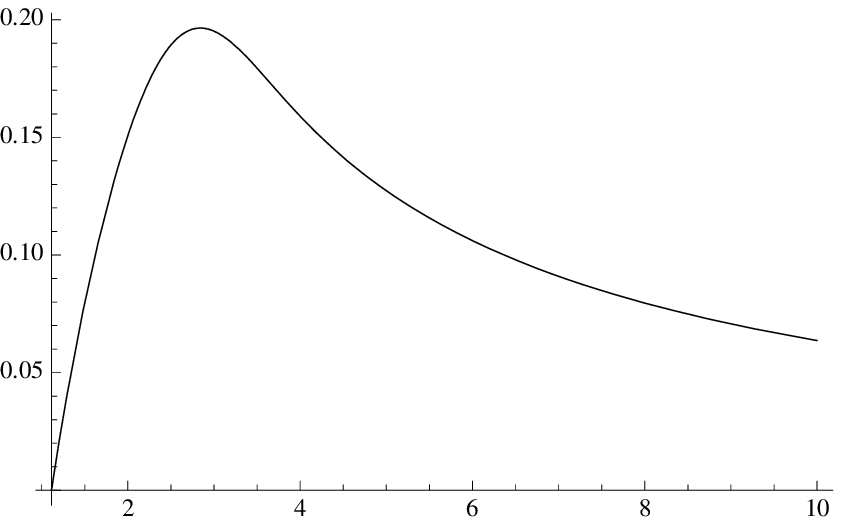}
\hspace{0.5cm}
\raisebox{3cm}{$\frac{\bar\beta_{10}^{-1}}{\bar m}$}
\epsfxsize=6.5cm
\epsfbox{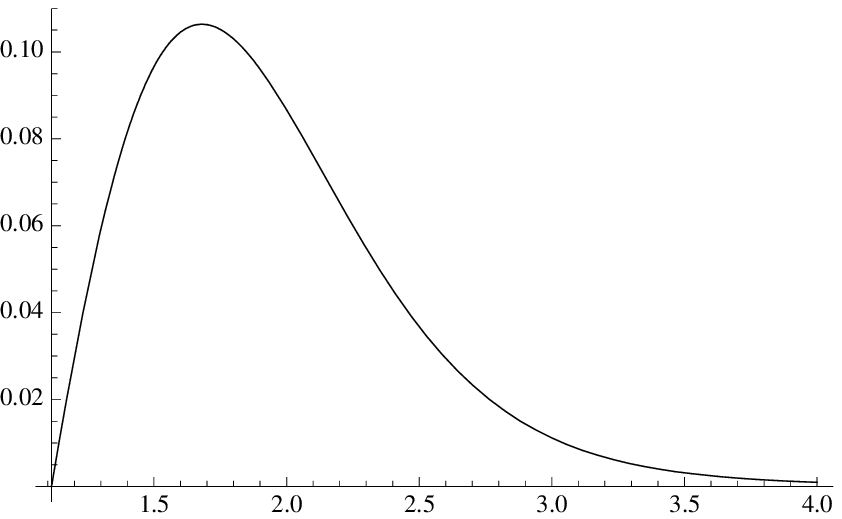}
\\
\hspace{5cm}
$\bar r_+$
\hspace{7cm}
$\bar r_+$
\caption{Temperature (left panel) and its ratio with proper mass (right panel)
{\em vs\/} horizon radius in $d=10$ (where $\bar r_+\ge 1.11$).
The maximum temperature $\bar\beta_{10}^{-1}\simeq 0.11\,\bar m$
is reached for $\bar r_+\simeq 1.68$.
\label{temp}}
\end{figure}

In Fig.~\ref{temp}, we then plot the mBH temperature $\bar\beta_{10}^{-1}$
(left panel) and its ratio with the proper mass $\bar m$ as functions of the horizon
radius $\bar r_+$ (right panel).
Since the latter ratio becomes smaller for lower $d$, it is clear that the mBH mass
always remains significantly larger than the typical energy of emitted quanta
(proportional to the temperature) and the canonical expression~\eqref{dMdt}
applies throughout for all values of $d$ (and all choices of $F_d$).
\par
\begin{figure}[b]
\centering
\raisebox{3cm}{$\bar{\mathcal{L}}_{10}$}
\epsfxsize=6.5cm
\epsfbox{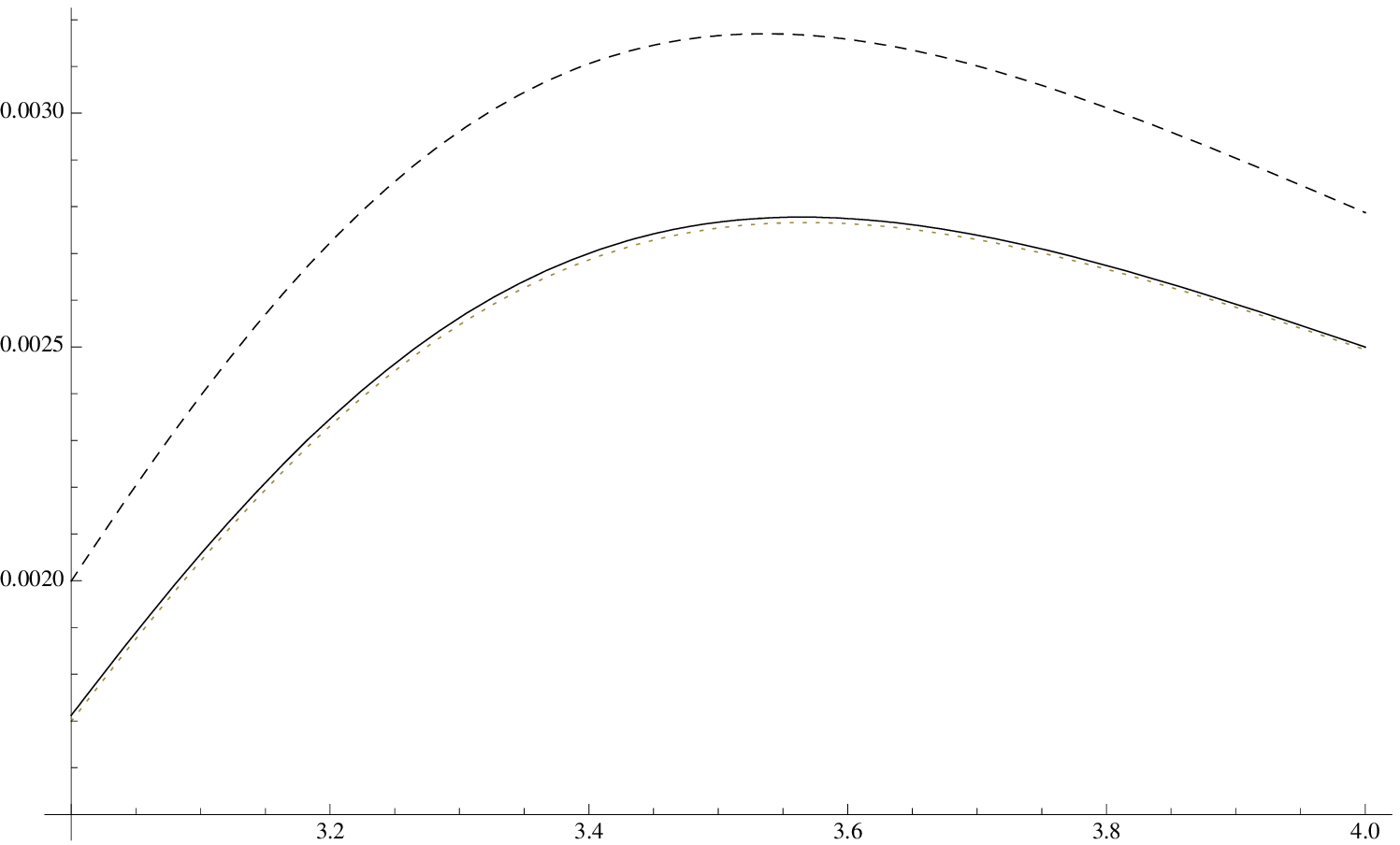}
\hspace{0.5cm}
\raisebox{3cm}{$\ $}
\epsfxsize=6.5cm
\epsfbox{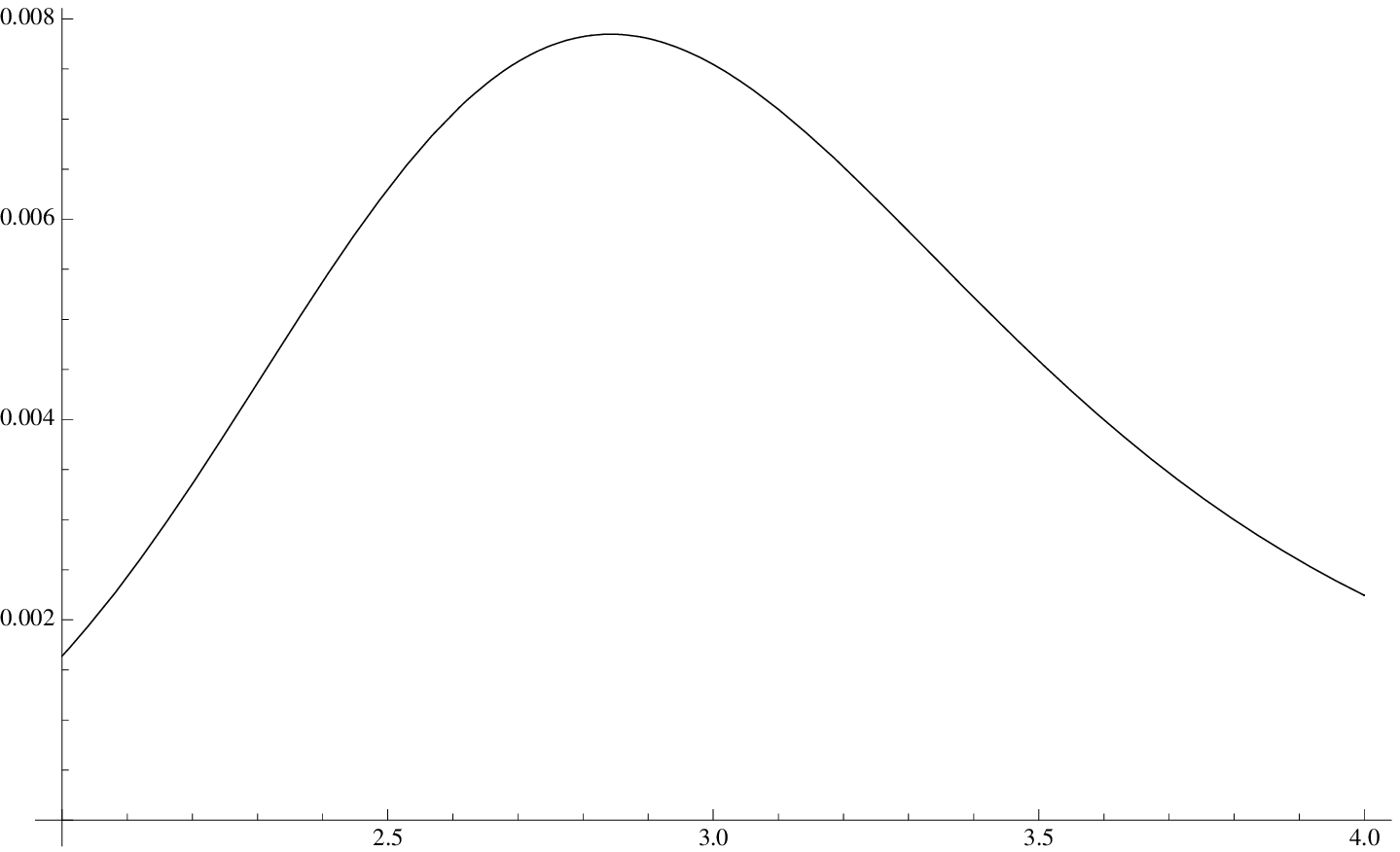}
\\
\hspace{5cm}
$\bar r_+$
\hspace{7cm}
$\bar r_+$
\caption{Left panel:
approximate mBH luminosity for $n_{\rm max}=0$ (dashed line), $2$
(solid line) and $10$ (dotted line).
Right panel: relative difference in luminosity for $n_{\rm max}=2$ and $n_{\rm max}=10$.
\label{dMerr}}
\end{figure}

Next, we need to determine how many terms in the sum in Eq.~\eqref{drdtx} we should
keep.
In Fig.~\ref{dMerr}, left panel, we show the dimensionless luminosity in $d=10$ with
$F_{10}=F_{10}^{\rm B}$ and $C=1$ for $n_{\rm max}=0$, $2$ and $10$ around the
peak, where discrepancies are the largest.
We can thus conclude that it is well sufficient to use $n_{\rm max}=2$
(the relative difference with respect to $n_{\rm max}=10$ does not depend on the choice
of $F_d$ and $C$ and is less than $8\cdot 10^{-3}$; see right panel).
For smaller values of $d$, this approximation actually becomes even better.
For example, for $d=5$ one obtains the smallest relative error of less than
$6\cdot 10^{-5}$.
\par
\begin{figure}[t]
\centering
\raisebox{3cm}{$\bar r_+$}
\epsfxsize=6.5cm
\epsfbox{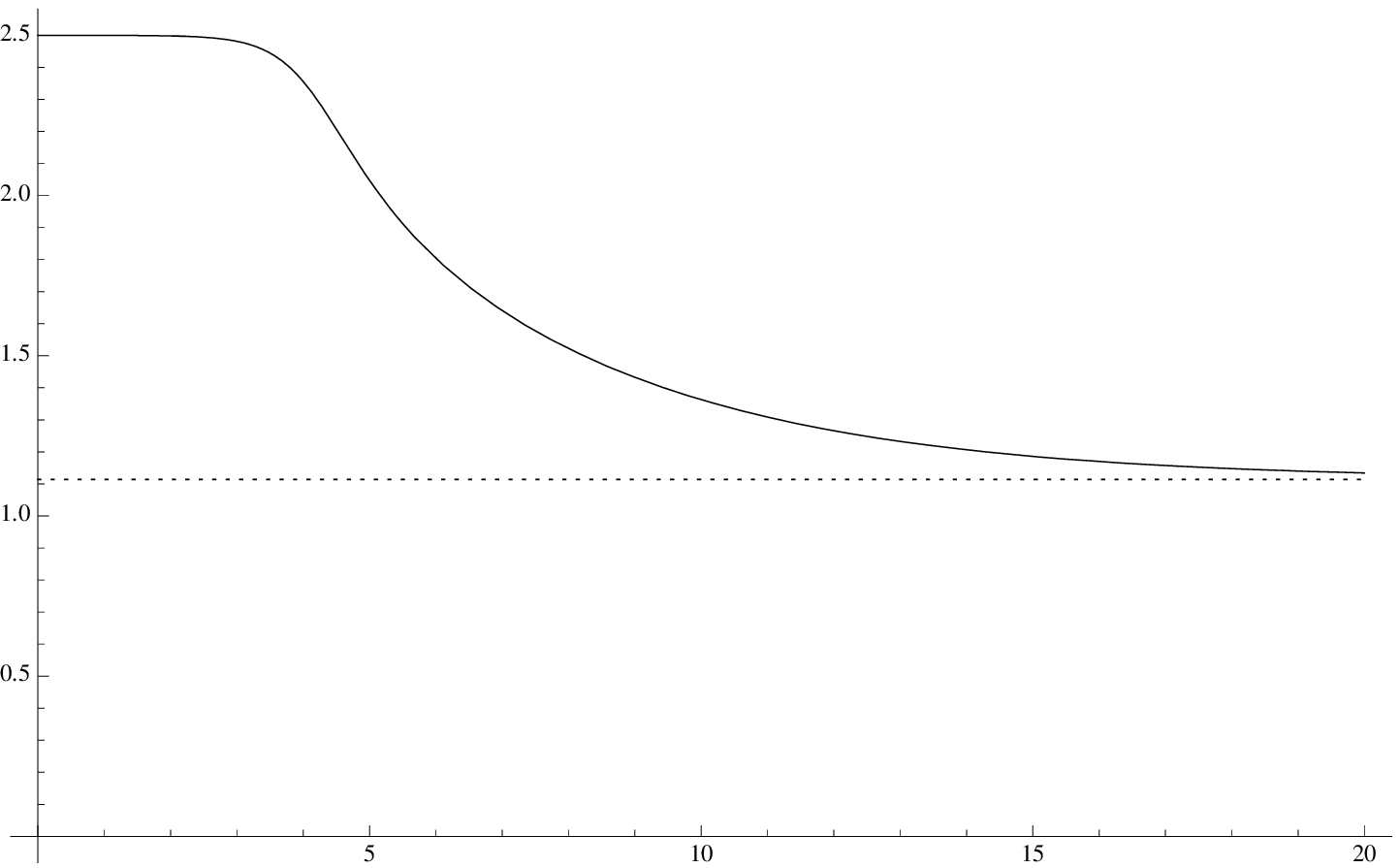}
\hspace{0.5cm}
\raisebox{3cm}{$\tilde m$}
\epsfxsize=6.5cm
\epsfbox{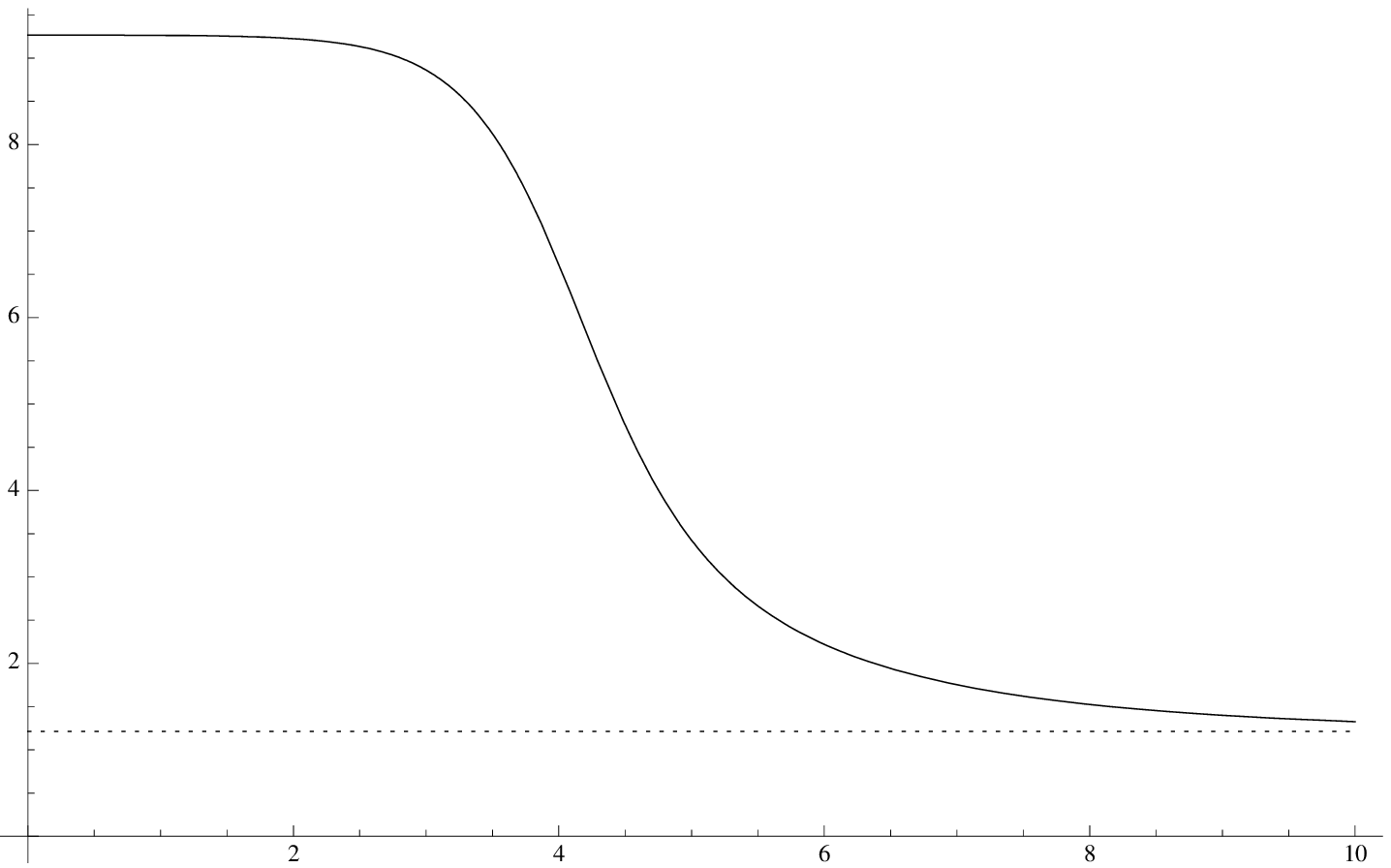}
\\
\hspace{5cm}
$\log_{10}(\bar t)$
\hspace{6cm}
$\log_{10}(\bar t)$
\caption{Left panel:
horizon radius for initial mBH mass $m\simeq 9.2\,$TeV (solid line)
and its minimum value (dotted line).
Right panel:
mBH proper mass for same initial condition (solid line) and remnant mass
$m_{\rm min}\simeq 1.2\,$TeV (dotted line).
\label{rtx}}
\end{figure}

From Eq.~\eqref{Madm} and~\eqref{m}, we find that an initial mBH proper mass
of about $10\,$TeV corresponds to an horizon radius $\bar r_+\simeq 2.5$ for $d=10$.
With this initial condition, Eq.~\eqref{drdtx} with $C=1$ and $F_{10}=F_{10}^{\rm B}$
yields the solution plotted in the left panel of Fig.~\ref{rtx}, from which the time-dependent
proper mass can be easily obtained (right panel).
The latter plot shows that the decay occurs through several stages:
it is initially rather slow (for $0\le \bar t\lesssim 300$),
subsequently becomes much faster (for $300\lesssim \bar t\lesssim 10^{6}$)
and finally approaches the remnant again slowly (for $\bar t\gtrsim 10^6$).
For our choice of $\ell_g=1\,$TeV$^{-1}\simeq 1.8\,\ell$, we obtain that
$\ell\simeq 0.6\,$TeV$^{-1}\simeq 3\cdot 10^{-28}\,$sec
and the estimated decay-time $t_{10}\simeq 10^{10}\,\ell\simeq 10^{-18}\,$sec is
practically instantaneous.
This result changes rather weakly with $d$, decreasing down to the minimum
$t_{5}\simeq 10^{6}\,\ell\simeq 10^{-22}\,$sec.
As we mentioned before, different choices of $F_d$ produce slightly different decay
curves, but do not yield longer decay-times.
For example, for $d=5$ and $F_{5}=F_{5}^{\rm b}$ in Eq.~\eqref{Fdb} one obtains the
shortest possible value for $C=1$, that is $t_5\simeq 10^4\,\ell\simeq 10^{-24}\,$sec.
For different values of $C$, all the decay-times scale roughly linearly.
Assuming $C\gtrsim 10^{-2}$ and considering the case $d=10$ with $F_{10}=F_{10}^{\rm B}$
thus yields the fairly conservative upper bound
\be
t_{\rm decay}\lesssim 10^{-16}\,{\rm sec}
\ ,
\label{td}
\ee
which is extremely short compared to the sensitivity of present detectors (on the order
of hundreds of picoseconds).
\section{Conclusions}
\setcounter{equation}{0}
We have estimated the decay-times of non-commutative mBH's that might be produced
at the LHC and found that they should evaporate nearly instantaneously (see the upper bound
in Eq.~\eqref{td}), much the same as is expected according to the ``canonical'' scenario
of Ref.~\cite{dimopoulos}.
We can consequently conclude that non-commutative mBHs would evaporate
within the detectors, since they can propagate at most a few nanometers away from
their point of production during the evaporation.
\par
The present results are similar to previous estimates of decay-times for usual
(``commutative'') mBH's in the ADD brane-world~\cite{add}, described according to
the microcanonical picture~\cite{mfd}, for which life-times were obtained on the order of
$10^{-17}\,$sec or shorter~\cite{ch02}.
Remarkably, our conclusions are also comparable with the recently estimated life-times of
mBH's derived from Generalised Uncertainty Principles in Ref.~\cite{scardigli}.
Actually, this is not surprising, since the dependence of the temperature on the mBH mass is
roughly similar in all of these cases~\cite{c07}.
Of course, the mBH's of Ref.~\cite{ch02} evaporate completely (thus reaching zero
temperature for vanishing mass) and their life-time is actually dominated by the latest
stages (when $m\lesssim 5\,$TeV), whereas the end-point of the process here, as well
as in Ref.~\cite{scardigli}, is a (presumably) stable remnant of mass
$m_{\rm min}\simeq 1\,$TeV and zero Hawking temperature.
\par
Let us conclude by mentioning that no accretion mechanism has been included in
our analysis of the evaporation, which is fully consistent given the very short decay-times.
What happens afterwards and the final fate of the remnant are still open questions.
\subsection*{Acknowledgments}
P.N.~is partially supported by a CSU Fresno International Activities Grant.
\end{document}